\documentclass[aps,showpacs,prd,twocolumn]{revtex4}%
\usepackage{graphicx}
\usepackage{amssymb}
\usepackage{amsmath}
\usepackage{color}
\usepackage{float}
\usepackage{ulem}
\usepackage{accents}
\usepackage{graphicx}
\usepackage{graphicx}
\usepackage{amsfonts}
\usepackage[colorlinks=true,pdfstartview=FitV,linkcolor=blue,citecolor=blue,urlcolor=blue,breaklinks=true]%
{hyperref}%
\providecommand{\U}[1]{\protect\rule{.1in}{.1in}}
\begin{document}
\title{Lorentz-violating contributions to the nuclear Schiff moment and nuclear EDM}
\author{Jonas B. Araujo}
\email{jonas.araujo88@gmail.com}
\author{Rodolfo Casana}
\email{rodolfo.casana@gmail.com}
\author{Manoel M. Ferreira Jr}
\email{manojr.ufma@gmail.com}
\affiliation{Departamento de F\'{\i}sica, Universidade Federal do Maranh\~{a}o, Campus
Universit\'{a}rio do Bacanga, S\~{a}o Lu\'{\i}s - MA, 65080-805 - Brazil}

\begin{abstract}
In the context of an atom endowed with nuclear electric dipole moment (EDM),
we consider the effects on the Schiff moment of $CPT$-even Lorentz-violating
(LV) terms that modify the Coulomb potential. First, we study the
modifications on the Schiff moment when the nucleus interacts with the
electronic cloud by means of a Coulomb potential altered only by the $P$-even
LV components. Next, by supposing the existence of an additional intrinsic LV
EDM generated by other LV sources, we assess the corrections to the Schiff
moment when the interaction nucleus-electrons runs mediated by a Coulomb
potential modified by both the $P$-odd and $P$-even LV components. We then use
known estimates and EDM measurements to discuss upper bounds on the new Schiff
moment components and the possibility of an intrisic nuclear EDM component
ascribed to LV effects.

\end{abstract}

\maketitle

\section{Introduction}

The possibility of detecting permanent electric dipole moments (EDMs) cannot
be underestimated since it is related to a few major problems in contemporary
Physics \cite{EDM1}, \cite{EDM2}, \cite{LeptonEDM}, \cite{Jungmann}. The EDM
interactions are $P$ (parity) odd and $T$ (time reversal) odd and,
consequently, $CP$ (charge conjugation $\times$ parity) odd (if the $CPT$
theorem holds). The breaking of $CP$ is one of the crucial Sakharov conditions
for the baryon asymmetry in the universe \cite{Sakharov}, and the CKM matrix
is insufficient to account for it alone, so that there is room for new Physics
or some as-yet-unknown phenomenology from other sectors of the standard model.
In the strong interactions, $CP$-violations are parametrized by the $\theta$
term, which is extremely small for as of yet unknown reasons. This poses the
strong $CP$-problem, whose solution via spontaneous breaking of the
Peccei-Quinn symmetry involves axions \cite{PecceiQuinn}. Although yet
undetected, axions could induce oscillating EDMs {\cite{AxionEDM}.}

The $P$ and $T$-odd nuclear forces could generate EDM by rendering charge
fluctuations over a finite-sized nucleus. To date, the best experimental upper
bound on a nuclear EDM is $|d(^{199}\text{Hg})|<7.4\times10^{-30}%
\ e\,\text{cm}$ \cite{expEDM}. According to the Schiff theorem \cite{Schiff},
in an atom with a pointlike nucleus and nonrelativistic electrons that
interact only electrostatically, the nuclear EDM is completely screened at
first order by the atom's electrons \cite{Schiff}, causing no Stark spectrum
shift. For a finite-sized nucleus, however, the first order screening is no
longer complete, there appearing the nuclear Schiff moment, whose interaction
with the electrons generates atomic EDM \cite{Schiff}, \cite{MSchiff1}%
,\cite{correctedSM2}. Such a nuclear EDM might yield an electric dipole moment
for the atom as a whole by a process that involves the mixing of electron wave
functions of opposite parity. The Schiff moment physics has been extensively
investigated, with several discussions and corrections having been performed
upon it. Experimental and theoretical proposals to verify considerable
enhancements to the octupole and Schiff moments in heavy nuclei have been
considered \cite{Dzuba, FlambaumSolo, Engel}, \cite{MSchiffRel2}. A proper
relativistic treatment of the electrons in an atom with a finite-sized nucleus
was considered in Ref. \cite{FlambaumIntegral}, with the generalization of the
Schiff moment and the evaluation of the local dipole moment (LDM)
incorporating the relativistic corrections. Numerical evaluations of the
Schiff moment \cite{Ban} and their relation to the atomic EDM magnitude (for a
few heavy atoms) are also known \cite{Dzuba2}. Further developments include,
for example, a more general form of the Schiff moment obtained via
calculations at the operator level \cite{correctedSM1}, the evaluation of
internal nucleon contribution to the Schiff moment \cite{Dmitriev}, the
enhancement of Schiff and octupole moments (by more than $2$ orders of
magnitude) in atoms with asymmetrically deformed nuclei with collective $P$
and $T$-odd electromagnetic interactions \cite{CollectiveMoments}, analysis of
the Schiff theorem in ions and molecules \cite{MSchiffRel1}, and other
important results \cite{More}.

Lorentz-violating (LV) theories have been under investigation since the 1990s
in different theoretical frameworks. In the standard model extension (SME)
\cite{Colladay}, many developments were performed in several sectors of
interactions, yielding tight constraints on the magnitude of the violation
coefficients \cite{Russell}. In the gauge electromagnetic sector, many studies
have scrutinized the effects of the $CPT$-odd \cite{CFJ} and $CPT$-even terms
\cite{KM}. The $CPT$-even gauge photon sector is modified by the tensor
$(K_{F})^{\alpha\beta\mu\nu}$ \cite{KM},\cite{Escobar}, whose $P$-odd and
$P$-even anisotropic repercussions on the {Coulomb potential} were properly
evaluated \cite{Bailey}, \cite{Cadu}, \cite{EPJC}. Nuclear systems have also
been a suitable environment to test and examine Lorentz violation, involving
detailed analysis of beta decay \cite{Betadecay} and nuclear spin/magnetic
dipole moment calculations \cite{Stadnik}. Recently, some implications of the
$P$-even coefficients, $\left(  k_{DE}\right)  ^{ij}$, on the {Coulomb
potential} were investigated in atoms with non-null nuclear quadrupole moments
$\left(  Q_{ij}\right)  $. Estimating the energy anisotropy yielded by the
interaction of the nuclear quadrupole moment with valence protons, $\delta
E=K\left(  k_{DE}\right)  ^{ij}Q_{ij}$, very tight bounds\textbf{\ }were set
on the LV parameters \cite{LVquad}.

LV theories were also applied to address the muon magnetic dipole moment
\cite{Vargas}, the neutron EDM \cite{Altarev}, one-loop contributions to
lepton EDM induced by the a LV fermion term \cite{Haghighat}, and by the
$CPT$-odd gauge coefficient \cite{Malta}, which has also been examined in
other respects \cite{Santos}. Some $CPT$-even coefficients, originally
belonging to the tensor $(K_{F})^{\alpha\beta\mu\nu}$, were also considered in
nonminimal couplings between fermions and photons, with the focus on the EDM
generation \cite{FredeNM1,Jonas1}, \cite{Ykost}. Nonminimal couplings
involving higher derivatives \cite{Kostelec1},\cite{Schreck} and
higher-dimension operators \cite{Reyes} were also considered. However, to
date, no study has been proposed to investigate the impact of a spacetime
anisotropy, stemming from Lorentz symmetry violation, on the EDM issues
connected to the Schiff theorem or the nuclear Schiff moment. In the present
paper, we evaluate the contributions that a Lorentz-violating anisotropic
Coulomb potential may induce to the atomic EDM shielding problem. More
specifically, we calculate how the modified Coulomb potential and an intrinsic
nuclear EDM, coming from the $CPT$-even Lorentz-violating tensor,
$(K_{F})^{\alpha\beta\mu\nu}$, yields corrections to the residual interaction
known as the Schiff moment.

\section{Schiff moment in a Lorentz-violating environment}

We consider an atomic nucleus whose charge density is $\rho(\boldsymbol{r}%
)=\rho_{0}(\boldsymbol{r})+\delta\rho(\boldsymbol{r})$, where $\rho_{0}(r)$
corresponds to the spherically symmetric part, normalized to unity, $\int%
\rho_{0}(r)d^{3}r=Ze,$ and the charge density fluctuations due to $P$ and
$T$-odd nuclear interactions are encoded in $\delta\rho(r)$, as described in
Refs. \cite{MSchiff1},\cite{CollectiveMoments}. A neutral atom has an
electronic cloud with $N=Z$ electrons; otherwise one has an ion of ($N-Z$)
charge. The nuclear EDM arises from the $P$ and $T$-odd interactions, being
given by
\begin{equation}
\boldsymbol{d}=Ze\int\boldsymbol{r}\delta\rho(\boldsymbol{r})d^{3}r.
\end{equation}
Now we regard an anisotropic Lorentz-violating {Coulomb potential} studied in
Ref. \cite{EPJC}, which yields the following potential:
\begin{equation}
A_{0}(r)={\frac{1}{4\pi}}\frac{q}{r}\left(  (1-n)+\kappa^{ij}\frac{r^{i}r^{j}%
}{2r^{2}}\right)  , \label{CoulombMod}%
\end{equation}
for a point particle of charge $q,$ where $\kappa^{ij}=$ $\left(
k_{e-}\right)  ^{ij}$ is a symmetric, parity-even and traceless tensor, and
$n=tr(k_{DE})/3$, with $\left(  k_{DE}\right)  ^{ij}=-2(K_{F})^{0i0j}$ (see
Ref. \cite{KM,Escobar}). Our goal is to investigate how this modified
potential could change the conclusions of Schiff's theorem \cite{Schiff} or
contribute to the Schiff moment. The starting point is writing out the
Hamiltonian for an atom in a region with an external field $\boldsymbol{E}%
_{0}$,%
\begin{equation}
H=K+V_{0(\text{LV})}+V+U_{(\text{LV})}+W, \label{Hfull1}%
\end{equation}
with $K$ representing the kinetic term,
\begin{equation}
K=-{\sum\limits_{i}^{N}}\frac{1}{2m_{e}}\frac{\partial^{2}}{\partial
\boldsymbol{R}_{i}^{2}}-\frac{1}{2M}\frac{\partial^{2}}{\partial
\boldsymbol{q}_{N}^{2}},
\end{equation}
while the electrostatic potential ($V)$ due to the external field is%
\begin{equation}
V=-\sum_{i}^{N}(-e\boldsymbol{R}_{i})\cdot\boldsymbol{E}_{0}-Ze\boldsymbol{q}%
_{N}\cdot\boldsymbol{E}_{0}\ .
\end{equation}
Here, $W=-\boldsymbol{d}\cdot\boldsymbol{E}_{0}$ is the interaction between
the nuclear EDM and the external field. Above, $\boldsymbol{R}_{i}$ and
$\boldsymbol{q}_{N}$ correspond to the $i$-th electron's and the nucleus
positions, respectively, while $\boldsymbol{r}$ is measured starting from the
nucleus' center. The anisotropic electrostatic potential between the atomic
components (electron$-$electron, electrons$-$nucleus), $V_{0(\text{LV})},$ now
receives contributions stemming from the Lorentz-violating potential
(\ref{CoulombMod}), that is,
\begin{align}
&  \left.  V_{0(\text{LV})}=e^{2}{\sum\limits_{i>j}^{N}}\left[  \frac
{(1-n)}{|\boldsymbol{R}_{i}-\boldsymbol{R}_{j}|}+\kappa^{kl}\frac
{(\boldsymbol{R}_{i}-\boldsymbol{R}_{j})^{k}(\boldsymbol{R}_{i}-\boldsymbol{R}%
_{j})^{l}}{2|\boldsymbol{R}_{i}-\boldsymbol{R}_{j}|^{3}}\right]  \right.
\nonumber\\
&  \left.  -Ze^{2}{\sum\limits_{i}^{N}}\int\rho_{0}(\boldsymbol{r})\left[
\frac{(1-n)}{|\boldsymbol{\tilde{R}}_{i}-\boldsymbol{r}|}+\kappa^{kl}%
\frac{(\boldsymbol{\tilde{R}}_{i}-\boldsymbol{r})^{k}(\boldsymbol{\tilde{R}%
}_{i}-\boldsymbol{r})^{l}}{2|\boldsymbol{\tilde{R}}_{i}-\boldsymbol{r}|^{3}%
}\right]  d^{3}r,\right. \nonumber
\end{align}
where $\boldsymbol{\tilde{R}}_{i}=(\boldsymbol{R}_{i}-\boldsymbol{q}%
_{N}\boldsymbol{)}$ gives the position of the $i$-th electron from the
nucleus. The same holds for the electrostatic potential, $U_{(\text{LV}%
)}=U_{(\text{is})}+U_{(\text{an})},$ which accounts for the interaction of the
$N$ electrons with the nuclear charge fluctuation, $\delta\rho(r),$ whose
isotropic and anisotropic LV interactions, $U_{(\text{is})},U_{(\text{an})}$,
are written as\textbf{\ }%
\begin{align}
U_{(\text{is})}  &  =-Ze^{2}\sum\limits_{i}^{N}\int\delta\rho(\boldsymbol{r}%
)(1-n)\frac{d^{3}r}{|\boldsymbol{\tilde{R}}_{i}-\boldsymbol{r}|},\label{Uis}\\
U_{(\text{an})}  &  =-Ze^{2}\sum_{i}^{N}\int\delta\rho(\boldsymbol{r})\left[
\frac{\kappa^{kl}}{2}\frac{(\boldsymbol{\tilde{R}}_{i}-\boldsymbol{r}%
)^{k}(\boldsymbol{\tilde{R}}_{i}-\boldsymbol{r})^{l}}{|\boldsymbol{\tilde{R}%
}_{i}-\boldsymbol{r}|^{3}}\right]  d^{3}r, \label{Uan}%
\end{align}
respectively. In order to investigate the possible contributions, we need to
expand $U_{(\text{LV})}$ in powers of $|\boldsymbol{\tilde{R}}_{i}|^{-1}$. In
this task, there appear terms such as
\begin{align}
\partial_{i}\partial_{j}\partial_{k}\left(  \frac{1}{\tilde{R}}\right)   &
=\left[  \partial_{i}\partial_{j}\partial_{k}-\frac{1}{5}(\delta_{ij}%
\partial_{k}+\delta_{ik}\partial_{j}+\delta_{jk}\partial_{i})\partial
^{2}\right]  \frac{1}{\tilde{R}}\nonumber\\
&  +\frac{1}{5}(\delta_{ij}\partial_{k}+\delta_{ik}\partial_{j}+\delta
_{jk}\partial_{i})\partial^{2}\frac{1}{\tilde{R}}\ . \label{OctupoleOp}%
\end{align}
In accordance with the literature \cite{MSchiff1} ,\cite{MSchiffRel2},
\cite{MSchiffRel1}, the term inside brackets in Eq. (\ref{OctupoleOp})
corresponds to an octupole moment operator that can be neglected. In addition,
the charge octupole,
\begin{equation}
o_{ijk}=r^{i}r^{j}r^{k}-\frac{r^{2}}{5}(\delta_{ij}r^{k}+\delta_{ik}%
r^{j}+\delta_{jk}r^{i}), \label{OctupoleCharge}%
\end{equation}
can also be used to manipulate terms appearing in the expansions
\cite{CollectiveMoments} and to further separate the ones in $o_{ijk}$. It
just so happens that the usual Schiff moment calculation yields the same
result for both procedures. This is not the case, however, when anisotropic
pieces are involved. First, we expand its isotropic part (the first piece),
$U_{(\text{is})},$ as
\begin{align}
U_{(\text{is)}}  &  \approx(1-n)\left\{  -e\boldsymbol{d}\cdot\sum_{i}%
^{N}\frac{\boldsymbol{\tilde{R}}_{i}}{|\boldsymbol{\tilde{R}}_{i}|^{3}}\right.
\nonumber\\
&  \left.  -4\pi\frac{Ze^{2}}{10}\int d^{3}r\delta\rho(\boldsymbol{r}%
)r^{2}\boldsymbol{r}\cdot\sum_{i}^{N}\nabla_{i}[\delta(\boldsymbol{\tilde{R}%
}_{i})]\right\}  .
\end{align}
In order to expand the anisotropic part, we need to write
$a=(\boldsymbol{\tilde{R}}_{i}-\boldsymbol{r})^{k}(\boldsymbol{\tilde{R}}%
_{i}-\boldsymbol{r})^{l}/|\boldsymbol{\tilde{R}}_{i}-\boldsymbol{r}|^{3},$ as
a Taylor series, that is,%
\begin{align}
a  &  \approx-r^{m}\partial_{m}\left[  \frac{(\boldsymbol{\tilde{R}}_{i}%
)^{k}(\boldsymbol{\tilde{R}}_{i})^{l}}{|\boldsymbol{\tilde{R}}_{i}|^{3}%
}\right]  +\frac{1}{6}r^{m}r^{p}r^{q}\times\nonumber\\
&  \left[  \partial_{k}\partial_{m}\partial_{p}\partial_{q}\left(  \frac
{1}{|\boldsymbol{\tilde{R}}_{i}|}\right)  (\boldsymbol{\tilde{R}}_{i}%
)^{l}+\partial_{p}\partial_{q}\partial_{k}\left(  \frac{1}{|\boldsymbol{\tilde
{R}}_{i}|}\right)  \delta_{ml}\right. \nonumber\\
&  \left.  +\partial_{m}\partial_{q}\partial_{k}\left(  \frac{1}%
{|\boldsymbol{\tilde{R}}_{i}|}\right)  \delta_{pl}+\partial_{m}\partial
_{p}\partial_{k}\left(  \frac{1}{|\boldsymbol{\tilde{R}}_{i}|}\right)
\delta_{ql}\right]  , \label{AnisotropicULV}%
\end{align}
where we omitted\textbf{\ }the zeroth and second order terms, for $\delta
\rho(\boldsymbol{r})$ is odd. We have also used the fact that
\begin{equation}
\frac{(\boldsymbol{\tilde{R}}_{i})^{k}(\boldsymbol{\tilde{R}}_{i})^{l}%
}{|\boldsymbol{\tilde{R}}_{i}|^{3}}=-\partial_{k}\left(  \frac{1}%
{|\boldsymbol{\tilde{R}}_{i}|}\right)  (\boldsymbol{\tilde{R}}_{i})^{l}.
\end{equation}
Note, still, that the term
\begin{equation}
r^{m}r^{p}r^{q}\partial_{k}\partial_{m}\partial_{p}\partial_{q}\left(
\frac{1}{|\boldsymbol{\tilde{R}}_{i}|}\right)  (\boldsymbol{\tilde{R}}%
_{i})^{l},
\end{equation}
in Eq. (\ref{AnisotropicULV}), can only be unambiguously rewritten by grouping
$r^{m}r^{p}r^{q}$ as in Eq. (\ref{OctupoleCharge}), in which we ignore the
charge octupole term, $o_{ijk},$ as usual. After this step, this ambiguity is
removed and we can use (\ref{OctupoleOp}) to address the remaining
derivatives. This procedure enables us to correctly expand the anisotropic
term, achieving the full expression%
\begin{align}
&  U_{(\text{LV})}\approx(1-n)\left\{  -e\boldsymbol{d}\cdot\sum_{i}^{N}%
\frac{\boldsymbol{\tilde{R}}_{i}}{|\boldsymbol{\tilde{R}}_{i}|^{3}}\right.
\nonumber\\
&  \left.  -4\pi\frac{Ze^{2}}{10}\int d^{3}r\delta\rho(\boldsymbol{r}%
)r^{2}\boldsymbol{r}\cdot\sum_{i}^{N}\nabla_{i}[\delta(\boldsymbol{\tilde{R}%
}_{i})]\right\} \nonumber\\
&  +e\sum_{i}^{N}\frac{(\kappa)^{kl}}{2}d^{m}\partial_{m}\left[
\frac{(\boldsymbol{\tilde{R}}_{i})^{k}(\boldsymbol{\tilde{R}}_{i})^{l}%
}{|\boldsymbol{\tilde{R}}_{i}|^{3}}\right] \nonumber\\
&  +4\pi\sum_{i}^{N}\int r^{2}\delta\rho(\boldsymbol{r})\frac{\kappa^{kl}}%
{2}\frac{Ze^{2}}{10}\times\nonumber\\
&  \left[  (\boldsymbol{\tilde{R}}_{i})^{l}r^{i}\partial_{i}\partial_{k}%
+\frac{2}{5}\delta_{kl}r^{i}\partial_{i}+\frac{9}{5}r^{l}\partial_{k}\right]
d^{3}r\delta(\boldsymbol{\tilde{R}}_{i}). \label{ULVexpansion}%
\end{align}
Now it is necessary to introduce the displacement operator%
\begin{equation}
Q_{\text{D}}=\frac{\boldsymbol{d}}{Ze}\cdot\frac{\partial}{\partial
\boldsymbol{q}_{N}}\ , \label{DisplacementQ}%
\end{equation}
which allows us to compute the commutator $[Q_{\text{D}},V_{0(\text{LV}%
)}]=\frac{\boldsymbol{d}}{Ze}\cdot\frac{\partial}{\partial\boldsymbol{q}_{N}%
}[V_{0(\text{LV})}].$ Deriving with respect to the coordinates $\boldsymbol{q}%
_{N}^{l}$ is equivalent to deriving in the components $\left(  \boldsymbol{R}%
_{i}\right)  ^{l}$, since $\partial_{l(\boldsymbol{q}_{N})}F(\boldsymbol{R}%
_{i}-\boldsymbol{q}_{N})=-\partial_{l(\boldsymbol{R}_{i})}F(\boldsymbol{R}%
_{i}-\boldsymbol{q}_{N}),$ or simply $\partial/\partial\boldsymbol{q}%
_{N}=-\partial/\partial\boldsymbol{R}_{i}$. Expanding the derivatives
$\partial V_{0(\text{LV})}/\partial\boldsymbol{R}_{i}$ (in a similar way to
the ones performed on $U_{(\text{LV})}$), making simplifications, the
following commutator is obtained:%
\begin{align}
&  [Q_{\text{D}},V_{0(\text{LV})}]=U_{(\text{LV})}-4\pi e\sum_{i}^{N}\frac
{1}{10}\times\nonumber\\
&  \left\{  Ze\int d^{3}r\delta\rho(\boldsymbol{r})r^{2}\left[  (1-n)r^{k}%
-\frac{9}{10}r^{l}\kappa^{kl}\right]  \right. \nonumber\\
&  \left.  -\frac{5}{3}\int r^{2}\rho_{0}(\boldsymbol{r})d^{3}r\left[
(1-n)d^{k}-\frac{2}{5}d^{l}\kappa^{kl}\right]  \right\}  \partial_{k}%
\delta(\boldsymbol{\tilde{R}}_{i}) \label{Commut}%
\end{align}
where $U_{(\text{LV})}$ is given by Eq. (\ref{ULVexpansion}) and $\partial
_{k}\delta(\boldsymbol{\tilde{R}}_{i})=\partial_{k(\boldsymbol{R}%
_{i}-\boldsymbol{q}_{N})}[\delta(\boldsymbol{R}_{i}-\boldsymbol{q}_{N})].$
This calculation implies that $[Q_{\text{D}},V_{0\text{(LV)}}]=U_{(\text{LV}%
)}-H_{\text{residual}}.$ On the other hand, $[Q_{\text{D}},V]=W$,
$[Q_{\text{D}},K]=0,$ so that the full Hamiltonian (\ref{Hfull1}) can be
written as%
\begin{equation}
H=H_{0(\text{LV})}+[Q_{\text{D}},H_{0(\text{LV})}]+H_{\text{residual}},
\label{Hfull2}%
\end{equation}
with $H_{0(\text{LV})}=K+V_{0\text{(LV)}}+V$ and $U_{(\text{LV})}%
+W=[Q_{\text{D}},H_{0(\text{LV})}]-H_{\text{residual}}$. Eq. (\ref{Hfull2})
implies the shielding of the pointlike dipole except for the residual
interaction, consequence of the finite nuclear size, which provides the atom
with EDM. In this case, the atomic EDM will be generated by a Schiff-like
moment that appears inside the residual interaction term, $H_{\text{residual}%
}=H_{\text{Schiff}(\text{LV})},$ read from Eq. (\ref{Commut}) as%
\begin{equation}
H_{\text{Schiff}(\text{LV})}=-4\pi e\mathbb{S}_{(1)}^{k}\sum_{i}^{N}%
\partial_{k}\delta(\boldsymbol{\tilde{R}}_{i}),
\end{equation}
where the modified Schiff moment,%
\begin{equation}
\mathbb{S}_{(1)}^{k}=(1-n)S^{k}-S_{(\text{LV1})}^{k}, \label{MSchiff}%
\end{equation}
is composed of two contributions,%
\begin{align}
S^{k}  &  =\frac{1}{10}\left(  Ze\int\delta\rho(\boldsymbol{r})r^{2}r^{k}%
d^{3}r-\frac{5}{3}d^{k}\int r^{2}\rho_{0}(\boldsymbol{r})d^{3}r\right)
,\label{usualSM}\\
S_{(\text{LV1})}^{k}  &  =\frac{\kappa^{kl}}{10}\left(  Ze\frac{9}{10}%
\int\delta\rho(\boldsymbol{r})r^{2}r^{l}d^{3}r-\frac{2}{3}d^{l}\int r^{2}%
\rho_{0}(\boldsymbol{r})d^{3}r\right)  . \label{SMLV}%
\end{align}
The usual Schiff moment, $S^{k}$, is induced by the $P$ and $T$-odd
interactions that generate the charge fluctuations $\delta\rho(r),$ while
$S_{(\text{LV1})}^{k}$ is the anisotropic piece associated with the LV tensor,
$\kappa^{kl}.$ The Lorentz violation also contributes by yielding the factor
$(1-n)$, which alters the weight of the usual moment ($S^{k})$ on the total
modified Schiff moment. Although the LV term containing $\kappa^{kl}$ does not
act as source of elementary dipole moment, $\boldsymbol{d}$, it creates a new
Schiff moment component, which impacts the atomic electric dipole moment in an
anisotropic manner. In writing $S_{(\text{LV1})}^{k}=\kappa^{kl}\tilde{S}%
^{l},$ with%
\begin{equation}
\tilde{S}^{l}=\frac{1}{10}\left[  Ze\frac{9}{10}\int\delta\rho(\boldsymbol{r}%
)r^{2}r^{l}d^{3}r-\frac{2}{3}d^{l}\int r^{2}\rho_{0}(\boldsymbol{r}%
)d^{3}r\right]  ,
\end{equation}
and comparing the expressions (\ref{usualSM}) and (\ref{SMLV}), one notes that
the LV kernel, $\tilde{S}^{l},$ has in principle the same order of magnitude
and the same direction as the usual Schiff moment, $S^{k}.$

A point worth mentioning is that the usual Schiff moment is aligned with the
nuclear spin $\left(  \mathbf{I}\right)  $, that is, $\boldsymbol{S}%
=S\mathbf{\hat{I}.}$ The same holds for $\mathbf{\tilde{S}}$. In contrast, the
Lorentz-violating piece (\ref{SMLV}) is no longer aligned with the spin, for
it is rotated by the matrix $\kappa^{kl},$ so that%
\begin{equation}
\mathbb{S}_{(1)}^{k}=S\mathbf{\hat{I}}^{k}+\tilde{S}(\kappa^{kl}%
\mathbf{\hat{I}}^{l})\mathbf{.} \label{SM2pieces}%
\end{equation}

We note the existence of an EDM associated with the LV piece, $S_{(\text{LV1}%
)}^{k},$ which we represent as $\left(  \mathbf{d}_{\text{LV}}\right)
^{k}=\alpha\kappa^{kl}\mathbf{d}^{l},$ being $\mathbf{d}^{l}$ the EDM
associated with the usual Schiff moment and $\alpha$ a constant. Concerning
the measurement respects, it is worth to discuss how the Schiff moment
components not aligned with the spin could become manifest in usual
experiments designed to detect nuclear EDM of atoms. In typical setups, an
atom endowed with spin $\mathbf{I,}$ magnetic moment $\boldsymbol{\mu=}%
\mu\mathbf{I}$ and EDM $\mathbf{d}=d\mathbf{I},$ {is placed in a region} with
an electric and a magnetic field, which cause a kind of Zeeman interaction,
$U=-(\mu\mathbf{B}+d\mathbf{E)\cdot I}$. {This interaction implies} precession
frequency around the $\boldsymbol{B}$ axis equal to $\omega_{i}=(\mu B+dE)/2$
(for spin 1/2 systems). {These experiments begin with parallel $\boldsymbol{E}%
$ and $\boldsymbol{B}$ fields, then} the electric field is {inverted},
modifying the precession frequency to $\omega_{f}=(\mu B-dE)/2$. The measured
{precession variation is a response due solely to the EDM}, $\Delta\omega=dE$.
In principle, the fact that the LV piece $S_{(\text{LV1})}^{k}$ is not
parallel to the spin does not avoid its detection, since the spin, even if
initially prepared in one particular axis, will precess around the magnetic
field direction. The point is that the LV EDM piece will yield precession
around the magnetic field as well, inasmuch as it also is written in terms of
the nuclear spin. It implies a variation on the frequency precession,
$\Delta\omega_{LV}=\left\vert \mathbf{d}_{LV}\right\vert E,$ after inverting
the electric field. The key experimental point is to find a way in which this
new contribution could be separated from the usual one.

\section{Intrinsic LV Nuclear EDM}

The $P$ and $T$-odd interactions act as possible generators of nuclear
intrinsic EDM in the atom. In a $CPT$-even scenario, the EDM interactions are
also $CP$-odd, however, they can be $CP$-even in a LV and $CPT$-odd framework.
It is worth supposing other sources (beyond the usual nuclear interactions)
that could yield intrinsic EDM to the nucleus, not necessarily associated to
the nuclear spin. This can be performed by Lorentz-violating $P$ and $T$-odd
terms belonging to the $CPT$-even and $CPT$-odd quark sector of the SME
\cite{Colladay}. It can be also triggered by dimension-five $CPT$-odd and
$CP$-even nonminimal interactions between quarks and the electromagnetic field
\cite{Pospelov}. Once the intrinsic nuclear LV EDM ($\boldsymbol{d}%
_{\text{ilv}}$) is generated, it is uniformly distributed over the nucleus and
interacts with the electronic cloud by means of a modified Coulomb potential,
as the one of Eq. (\ref{CoulombMod}). Now, we examine this situation,
rewriting $U_{(\text{is})}$ and $U_{(\text{an})}$ in Eqs. (\ref{Uis}),
(\ref{Uan}), as
\begin{align}
U_{(\text{is})}^{\prime}  &  =-e(1-n)\boldsymbol{d}_{\text{ilv}}\cdot\sum
_{i}^{N}\frac{\boldsymbol{\tilde{R}}_{i}}{|\boldsymbol{\tilde{R}}_{i}|^{3}%
}\nonumber\\
&  -Ze^{2}\sum\limits_{i}^{N}\int\delta\rho(\boldsymbol{r})(1-n)\frac{d^{3}%
r}{|\boldsymbol{\tilde{R}}_{i}-\boldsymbol{r}|},\\
U_{(\text{an})}^{\prime}  &  =e\frac{\kappa^{kl}}{2}\sum_{i}^{N}\left[
-3\frac{(\boldsymbol{\tilde{R}}_{i})^{k}(\boldsymbol{\tilde{R}}_{i}%
)^{l}\boldsymbol{d}_{\text{ilv}}\cdot\boldsymbol{\tilde{R}}_{i}}%
{|\boldsymbol{\tilde{R}}_{i}|^{5}}+2\frac{\boldsymbol{d}_{\text{ilv}}%
\cdot\boldsymbol{\tilde{R}}_{i}}{|\boldsymbol{\tilde{R}}_{i}|^{3}}\right]
\nonumber\\
&  -Ze^{2}\sum_{i}^{N}\int\delta\rho(\boldsymbol{r})\left[  \frac{\kappa^{kl}%
}{2}\frac{(\boldsymbol{\tilde{R}}_{i}-\boldsymbol{r})^{k}(\boldsymbol{\tilde
{R}}_{i}-\boldsymbol{r})^{l}}{|\boldsymbol{\tilde{R}}_{i}-\boldsymbol{r}|^{3}%
}\right]  d^{3}r.
\end{align}
The new potential $\mathbb{U}^{\prime}=U_{(\text{is})}^{\prime}+U_{(\text{an}%
)}^{\prime}$ is expanded as before, yielding
\begin{align}
&  U_{(\text{LV2})}\approx(1-n)\left\{  -e(\boldsymbol{d}_{\text{ilv}%
}+\boldsymbol{d}_{\delta})\cdot\sum_{i}^{N}\frac{\boldsymbol{\tilde{R}}_{i}%
}{|\boldsymbol{\tilde{R}}_{i}|^{3}}\right. \nonumber\\
&  \left.  -4\pi\frac{Ze^{2}}{10}\int d^{3}r\delta\rho(\boldsymbol{r}%
)r^{2}\boldsymbol{r}\cdot\sum_{i}^{N}\nabla_{i}[\delta(\boldsymbol{\tilde{R}%
}_{i})]\right\} \nonumber\\
&  +e\sum_{i}^{N}\frac{\kappa^{kl}}{2}(\boldsymbol{d}_{\text{int}%
}+\boldsymbol{d}_{\delta})^{m}\partial_{m}\left[  \frac{(\boldsymbol{\tilde
{R}}_{i})^{k}(\boldsymbol{\tilde{R}}_{i})^{l}}{|\boldsymbol{\tilde{R}}%
_{i}|^{3}}\right] \nonumber\\
&  +4\pi\sum_{i}^{N}\int r^{2}\delta\rho(\boldsymbol{r})\frac{\kappa^{kl}}%
{2}\frac{Ze^{2}}{10}\times\nonumber\\
&  \left[  (\boldsymbol{\tilde{R}}_{i})^{l}r^{i}\partial_{i}\partial_{k}%
+\frac{2}{5}\delta_{kl}r^{i}\partial_{i}+\frac{9}{5}r^{l}\partial_{k}\right]
d^{3}r\delta(\boldsymbol{\tilde{R}}_{i}),
\end{align}
in which $\boldsymbol{d}_{\delta}$ stands for the EDM generated by the charge
fluctuations. The displacement operator (\ref{DisplacementQ}) is now rewritten
as
\begin{equation}
Q_{\text{D}}=\frac{(\boldsymbol{d}_{\text{ilv}}+\boldsymbol{d}_{\delta})}%
{Ze}\cdot\frac{\partial}{\partial\boldsymbol{q}_{N}},
\end{equation}
leading to the following commutator
\begin{align}
&  [Q_{\text{D}},V_{0(\text{LV2})}]=U_{(\text{LV2})}-4\pi e\sum_{i}^{N}%
\frac{1}{10}\times\nonumber\\
&  \left\{  Ze\int d^{3}r\delta\rho(\boldsymbol{r})r^{2}\left[  (1-n)r^{k}%
-\frac{9}{10}r^{l}\kappa^{kl}\right]  \right. \nonumber\\
&  -\frac{5}{3}\int r^{2}\rho_{0}(\boldsymbol{r})d^{3}%
r\bigg[(1-n)(\boldsymbol{d}_{\text{ilv}}+\boldsymbol{d}_{\delta}%
)^{k}\nonumber\\
&  \left.  -\frac{2}{5}(\boldsymbol{d}_{\text{ilv}}+\boldsymbol{d}_{\delta
})^{p}\kappa^{kp}\bigg]\right\}  \partial_{k}\delta(\boldsymbol{\tilde{R}}%
_{i}).
\end{align}
In comparison with Eq. (\ref{Commut}), one notes that the total EDM is
modified, $\boldsymbol{d}\rightarrow\boldsymbol{d}_{\text{ilv}}+\boldsymbol{d}%
_{\delta}$. Furthermore, the commutator $[Q_{\text{D}},V]$ still yields $W$,
so that the Schiff shielding\textbf{\ }is preserved at first order, and the
residual interaction keeps its form, $H_{\text{Schiff}(\text{LV})}=-4\pi
e\mathbb{S}_{(2)}^{k}\sum\limits_{i}^{N}\partial_{k}\delta(\boldsymbol{\tilde
{R}}_{i}),$ with%
\begin{equation}
\mathbb{S}_{(2)}^{k}=(1-n)S_{(2)}^{k}+S_{(\text{LV2})}^{k}, \label{MSchiff2}%
\end{equation}
being the modified Schiff moment, which now receives contributions from
$\boldsymbol{d}_{\text{ilv}}$ and $\boldsymbol{d}_{\delta}$, written in two
pieces,%
\begin{align}
S_{(2)}^{k}  &  =\frac{1}{10}\left[  Ze\int\delta\rho(\boldsymbol{r}%
)r^{2}r^{k}d^{3}r\right. \nonumber\\
&  \left.  -\frac{5}{3}(\boldsymbol{d}_{\text{ilv}}+\boldsymbol{d}_{\delta
})^{k}\int r^{2}\rho_{0}(\boldsymbol{r})d^{3}r\right]  ,\\
S_{(\text{LV2})}^{k}  &  =-\frac{\kappa^{kl}}{10}\left\{  Ze\frac{9}{10}%
\int\delta\rho(\boldsymbol{r})r^{2}r^{l}d^{3}r\right. \nonumber\\
&  \left.  -\frac{2}{3}(\boldsymbol{d}_{\text{ilv}}+\boldsymbol{d}_{\delta
})^{l}\int r^{2}\rho_{0}(\boldsymbol{r})d^{3}r\right\}  . \label{SMLV2}%
\end{align}
We can rewrite (\ref{MSchiff2}) at first order in terms of the usual Schiff
moment
\begin{equation}
\mathbb{S}_{(2)}^{k}=(1-n)S^{k}-{\frac{1}{6}(\boldsymbol{d}_{\text{ilv}}%
)^{k}\int r^{2}\rho_{0}(\boldsymbol{r})d^{3}r}+S_{(\text{LV1})}^{k},
\label{MSchiff2B}%
\end{equation}
where $S^{k}$ is the usual Schiff moment (\ref{usualSM}). Thus, in the
presence of an intrinsic EDM, the Schiff theorem is kept unharmed, with the
Schiff moment receiving a new contribution stemming from the intrinsic EDM.
Note that the modified Schiff moment (\ref{MSchiff2B}) generally does not
point in the same direction as the nuclear spin $\mathbf{I}$, since the
intrinsic moment ${\boldsymbol{d}_{\text{ilv}}}$ carries LV \textquotedblleft
rotations\textquotedblright. At first order, one can neglect $\boldsymbol{d}%
_{\text{ilv}}$ inside $S_{(\text{LV2})}^{k},$ given that $\boldsymbol{d}%
_{\text{ilv}}$ should also depend on LV coefficients, as commented in
Conclusions. {For this reason}, {the} LV Schiff moment (\ref{SMLV2}) becomes
equal {to the one} in Eq. (\ref{SMLV}).

In the following, we consider the case in which the total nuclear EDM,
$\boldsymbol{d}_{\text{ilv}}+\boldsymbol{d}_{\delta},$ interacts with the
electronic cloud by a Coulomb potential altered by $P$-odd coefficients
belonging to the tensor $(K_{F})^{\alpha\beta\mu\nu},$ as $\kappa^{j}=\frac
{1}{2}\epsilon^{jpq}\left(  \kappa_{DB}\right)  ^{pq},$ with $\left(
\kappa_{DB}\right)  ^{jk}=\epsilon^{kpq}\left(  K_{F}\right)  ^{0jpq}$
\cite{KM}. In Ref. \cite{Cadu}, the induced modifications to the {Coulomb
potential} were evaluated at{\ second order in $\kappa^{i}$: \ }$A_{0}\left(
\mathbf{r}\right)  =(q/4\pi)\left\{  (1+c_{\kappa})/r-(\boldsymbol{\kappa
}\cdot\mathbf{r})^{2}/2r^{3}\right\}  $, where $c_{\kappa}=-\mathbf{\kappa
}^{2}/2$, in which the anisotropy appears at second order. Following the same
steps, the modified Schiff moment is $\mathbb{S}_{(3)}^{k}=(1+c_{\kappa
})S_{(3)}^{k}-S_{(\text{LV3})}^{k},$ which now receives contributions from
$\boldsymbol{d}_{\text{ilv}}$ and $\boldsymbol{d}_{\delta}$, written in two
pieces,%
\begin{align}
S_{(3)}^{k}  &  =\frac{1}{10}\left[  Ze\int\delta\rho(\boldsymbol{r}%
)r^{2}r^{k}d^{3}r\right. \nonumber\\
&  \left.  -\frac{5}{3}(\boldsymbol{d}_{\text{ilv}}+\boldsymbol{d}_{\delta
})^{k}\int r^{2}\rho_{0}(\boldsymbol{r})d^{3}r\right]  ,\\
S_{(\text{LV3})}^{k}  &  =-\frac{\kappa^{k}\kappa^{l}}{10}\left\{  Ze\frac
{9}{10}\int\delta\rho(\boldsymbol{r})r^{2}r^{l}d^{3}r\right. \nonumber\\
&  \left.  -\frac{2}{3}(\boldsymbol{d}_{\text{ilv}}+\boldsymbol{d}_{\delta
})^{l}\int r^{2}\rho_{0}(\boldsymbol{r})d^{3}r\right\}  ,
\end{align}
which, at first order in the LV parameter ($\kappa^{l})$, is:%
\begin{equation}
\mathbb{S}_{(3)}^{k}=S^{k}-{\frac{1}{6}(\boldsymbol{d}_{\text{ilv}})^{k}\int
r^{2}\rho_{0}(\boldsymbol{r})d^{3}r}. \label{MSchiff3B}%
\end{equation}

\section{Conclusions and final remarks}

We have discussed the repercussions of LV terms belonging to the
electromagnetic $CPT$-even sector of the SME to the nuclear Schiff moment and
also the effect of an additional intrinsic LV nuclear EDM, $\boldsymbol{d}%
_{\text{ilv}}$. The achieved modified LV Schiff moments (\ref{MSchiff}),
(\ref{MSchiff2B}), (\ref{MSchiff3B}) do not point in the same direction as the
nuclear spin, as it occurs with the usual Schiff moment. Thus, it is important
to examine how these LV components can be detected, regarding the small
magnitude of all EDM pieces. In typical experiments to measure EDM, the atom
spin $\mathbf{I}$ is prepared in a given direction, and then it interacts with
controlled electric and magnetic fields, as already mentioned. A way to
isolate the LV contribution could be preparing the nuclear spin state to point
in the same direction as the magnetic field axis, $\mathbf{I}=\left\vert
\mathbf{I}\right\vert \hat{z}$, with $|s_{z}\ \pm\rangle$. In this
{configuration}, only the LV EDM component will yield torque and precession.
The $x$-component of the Schiff moment (\ref{SM2pieces}), therefore, is given
entirely in terms of $S_{(\text{LV1})}^{k},$ that is,
\begin{equation}
\mathbb{S}_{(1)}^{x}=\tilde{S}(\kappa^{13}\mathbf{\hat{I}}^{z}),
\end{equation}
so that the precession motion will be caused only by the $\kappa^{13}\tilde
{S}^{3}$ component. In such a situation, usual EDM experiments can be used to
constrain the magnitude of LV Schiff moment. Using recent calculations and
experimental data, we can set bounds on the LV Schiff moment components,
observing that $\tilde{S}$ and $S$ have the same order of magnitude. In Ref.
\cite{Dzuba2}, the EDM of a few atoms was estimated in terms of their Schiff
moments. Among them, the $^{199}\text{Hg}$ atom is the one which possesses the
most precise EDM measurement. According to those calculations, the magnitude
of the EDM is $d(^{199}\text{Hg})=-2.8\times10^{-17}\left(  S/e\ \text{fm}%
^{3}\right)  \ e\,\text{cm}$. The experimental data of Ref. \cite{expEDM} for
the $^{199}\text{Hg}$ EDM is, $|d(^{199}\text{Hg})|<3.1\times10^{-29}%
\ e\,\text{cm}$. In a setup as proposed above, the following upper bound is
obtained:%
\begin{equation}
|\langle\mathbb{S}_{(1)}^{x}\rangle_{z\pm}|=|\tilde{S}\kappa^{13}%
|\ \lessapprox2.6\times10^{-12}\ e\,\text{fm}^{3}, \label{bound13}%
\end{equation}
It is also relevant to point out that these bounds are set in the Earth's
reference frame (RF), where the experimental apparatus is located, and that
the LV coefficients are not constant in it. In order to write these results in
terms of the coefficients measured in a Sun-based RF, where these coefficients
are approximately constant, it is necessary to perform the sidereal analysis,
i.e., translate the bounds between these RFs. We will consider, as in the
literature \cite{Jonas1,Sideral}, the Earth-based Lab's RF at the colatitude
$\chi$, rotating around the Earth's axis with angular velocity $\Omega
=2\pi/23$h$\,56\text{s}$. For experiments up to a few weeks long, the
transformation law for a rank-2 tensor is $A_{ij}^{\text{(Lab)}}%
=\mathcal{R}_{ik}\mathcal{R}_{jl}A_{kl}^{\text{(Sun)}},$ with $\mathcal{R}%
_{ij}$ representing merely a spatial rotation
\begin{equation}
\mathcal{R}_{ij}=%
\begin{pmatrix}
\cos\chi\cos\Omega t & \cos\chi\sin\Omega t & -\sin\chi\\
-\sin\Omega t & \cos\Omega t & 0\\
\sin\chi\cos\Omega t & \sin\chi\sin\Omega t & \cos\chi
\end{pmatrix}
.
\end{equation}
The Earth-based RF has axis $x$, $y$ and $z$, while the Sun-based RF has $X$,
$Y$ and $Z$ as axis. Hence, $A_{ij}^{\text{(Lab)}}\equiv A_{ij}^{(x,y,z)}$ and
$A_{ij}^{\text{(Sun)}}\equiv A_{ij}^{(X,Y,Z)}$. Furthermore, by definition,
the $z$ axis matches the direction of the Earth's rotation axis and the $x$
axis points from the Earth's center to the Sun on the vernal equinox in 2000 -
for more details, see Refs. \cite{Sideral}.

According to the transformation law mentioned above, the time-averaged bound
(\ref{bound13}) is $\langle(\kappa)^{zx}\rangle^{(\text{Lab})}=\frac{1}%
{2}(\sin\chi\cos\chi)[(\kappa^{XX}+\kappa^{YY}-2\kappa^{ZZ})^{\text{Sun}}\ ]$.
Given that Tr($\kappa)=0,$ it holds that $\langle(\kappa)^{zx}\rangle
^{(\text{Lab})}=-\frac{3}{2}(\sin\chi\cos\chi)[(\kappa^{ZZ})^{(\text{Sun}%
)}\ ]$, so that the upper bound (\ref{bound13}) now constrains one of the
diagonal elements of the $\kappa$ matrix (in the Sun's RF),
\begin{equation}
\left\vert \tilde{S}\cos\chi\sin\chi\right\vert |\kappa^{ZZ}|\ \lessapprox
1.7\times10^{-13}\ e\,\text{fm}^{3}, \label{bound13B}%
\end{equation}
As one considers that $\tilde{S}$ has the same order of magnitude of the usual
Schiff moment, $\tilde{S}\sim10^{-12}\ e\,$fm$^{3}$, the bounds (\ref{bound13}%
) and (\ref{bound13B}) allow for major LV coefficients acting on the nucleus
physics, $|\kappa^{ZZ}|\ \lesssim0.17$, although without enhancing the atomic
EDM {- a possibility that certainly deserves more investigation}.

Such a scenario is also interesting if one takes into account the presence of
the intrinsic LV nuclear EDM, $\boldsymbol{d}_{\text{ilv}}$, stemming from LV
interactions. The estimates in the literature for the Schiff moment
\cite{Dzuba2} do not account for the last term in Eq. (\ref{MSchiff2B}) nor
Eq. (\ref{MSchiff3B}), which could amplify/modify the total Schiff moment.
First, it is necessary to make presumptions about the structure of this LV EDM
piece, such as it not being parallel to the nuclear spin $\mathbf{I}$. A first
possibility is to consider that $\boldsymbol{d}_{\text{ilv}}$ points in a
specific fixed direction in spacetime, given by the LV background under
consideration, without relation to the nuclear spin,\textbf{ }such as it
occurs in some LV nonminimal coupling systems, in which the LV background is
coupled directly to the field strength \cite{NMC}. In this case, the intrinsic
$\boldsymbol{d}_{\text{ilv}}$ would not cause spin precession around the
magnetic field, since it would not be written in terms of the spin
$\mathbf{I}$. {Furthermore}, the implied torque associated to the fixed
direction of $\boldsymbol{d}_{\text{ilv}}$ in spacetime could yield
dissipation of an initial precession. In this case, the rate of the
dissipation or change in the precession frequency would work as a channel for
constraining the\textbf{ }$\mathbf{d}_{\text{ilv}}$\textbf{ }magnitude.
\textbf{\ }It is worth mentioning that such a spin-independent EDM (or MDM) in
the direction of background does not appear in any nonminimal coupling, as the
$CPT$-even nonminimal couplings \cite{FredeNM1,Jonas1}, for instance. As a
second possibility, the EDM $d_{\text{ilv}}$ can also emerge as the result of
the presence of LV coefficients in the nuclear interactions, properly coupled
to the nuclear fields (instead of electromagnetic field). In this case, it is
expected that $\boldsymbol{d}_{\text{ilv}}$ should depend on both the spin
$\mathbf{I}$ and background fields directions, so that the intrinsic LV
nuclear EDM, $\boldsymbol{d}_{\text{ilv}},$ could also imply precession, being
subject to constraining by the procedure that led to the bound (\ref{bound13}).

Considering the possibility of separating the repercussions of parity-odd and
parity-even LV coefficients, we could choose to work with the Schiff moment
(\ref{MSchiff3B}) in order to focus on the $\boldsymbol{d}_{\text{ilv}}$
effects. Taking the nuclear spin aligned with the magnetic field in the
$z$-axis, the usual Schiff moment does need to be considered for precession
respects, so that the effective Schiff moment $S_{(3)}^{k}=(d_{\text{ilv}}%
^{k}/6)\int r^{2}\rho_{0}(r)d^{3}r$. Taking ${\boldsymbol{d}_{\text{ilv}}}$
and ${\boldsymbol{d}}_{\delta}$ proportional but not parallel, $\left\vert
{\boldsymbol{d}_{\text{ilv}}}\right\vert {=\beta}\left\vert {\boldsymbol{d}%
}_{\delta}\right\vert $, for a constant $\beta$, it holds {that}
$\boldsymbol{d}_{\text{ilv}}^{k}\int r^{2}\rho_{0}(\boldsymbol{r})d^{3}%
r=\beta\eta\tilde{S}\hat{k}.$ Here, it was used that the piece $\left\vert
\boldsymbol{d}_{\delta}\right\vert \int r^{2}\rho_{0}(\boldsymbol{r})d^{3}r$
has the same order of magnitude of the Schiff moment $\tilde{S}^{k},$
$\left\vert \boldsymbol{d}_{\delta}\right\vert \int r^{2}\rho_{0}%
(\boldsymbol{r})d^{3}r\sim\eta\tilde{S}.$ With {this} hypothesis, we obtain
$\mathbb{S}_{(3)}^{k}=\beta\tilde{S}\boldsymbol{\hat{d}}_{\text{ilv}}^{k}/6,$
with ($\eta\sim1).$ In this case, if the vector ${\boldsymbol{d}_{\text{ilv}}%
}$ has an $x$-component, it turns out that $\mathbb{S}_{(3)}^{x}=\beta
\tilde{S}/6.$ Following the route that led to bound (\ref{bound13}), one
achieves $\beta/6\lesssim0.7$ or $\beta\lesssim4.$ The attainment of a non
null $\beta,$ compatible with the current experimental measurements, can
indicate the existence of nuclear intrinsic EDM (due to LV effects only),
$\boldsymbol{d}_{\text{ilv}},$ and sensitive Lorentz violation in nuclear
systems. This is a new issue\textbf{ }that remains to be properly investigated
and involves the uncovering of the theoretical structure of $\boldsymbol{d}%
_{\text{ilv}}$.

\begin{acknowledgments}
The authors are grateful to CNPq, CAPES and FAPEMA (Brazilian research
agencies) for the financial support.
\end{acknowledgments}

\end{document}